\documentclass[twocolumn,showpacs,preprintnumbers,superscriptaddress,amsmath,amssymb]{revtex4-1}

\usepackage{epsfig}
\usepackage{graphicx}
\usepackage{dcolumn}
\usepackage{bm}
\usepackage{times}
\usepackage{xcolor}

\begin{document}


\title{Accurate ranking of influential
spreaders in networks based on dynamically asymmetric link-impact}

\author{Ying Liu}\email{shinningliu@163.com}
\affiliation{School of Computer Science, Southwest Petroleum University, Chengdu 610500, China}
\affiliation{Web Sciences Center, University of Electronic
Science and Technology of China, Chengdu 610054, China}

\author{Ming Tang} \email{tangminghuang521@hotmail.com}
\affiliation{Web Sciences Center, University of Electronic
Science and Technology of China, Chengdu 610054, China}
\affiliation{School of Information Science Technology, East China Normal University, Shanghai 200241, China}

\author{Younghae Do}
\affiliation{Department of Mathematics, Kyungpook National
University, Daegu 702-701, South Korea}

\author{Pak Ming Hui}
\affiliation{Department of Physics, Chinese University
of Hong Kong, Shatin, Hong Kong SAR, China}

\date{\today}

\begin{abstract}

We propose an efficient and accurate measure for ranking spreaders
and identifying the influential ones in spreading processes in
networks. While the edges determine the connections among the
nodes, their specific role in spreading should be considered
explicitly. An edge connecting nodes $i$ and $j$ may differ in its
importance for spreading from $i$ to $j$ and from $j$ to $i$. The
key issue is whether node $j$, after infected by $i$ through the
edge, would reach out to other nodes that $i$ itself could not
reach directly. It becomes necessary to invoke two unequal
weights $w_{ij}$ and $w_{ji}$ characterizing the importance of an
edge according to the neighborhoods of nodes $i$ and $j$. The
total asymmetric directional weights originating from a node leads
to a novel measure $s_{i}$ which quantifies the impact of the node
in spreading processes. A $s$-shell decomposition scheme further
assigns a $s$-shell index or weighted coreness to the nodes. The
effectiveness and accuracy of rankings based on $s_{i}$ and the
weighted coreness are demonstrated by applying them to nine
real-world networks. Results show that they generally outperform
rankings based on the nodes' degree and $k$-shell index, while
maintaining a low computational complexity. Our work represents a
crucial step towards understanding and controlling the spread of
diseases, rumors, information, trends, and innovations in
networks.

\end{abstract}

\maketitle

\section{Introduction} \label{sec:intro}
The structural properties of complex networks and the intricate
interplay between the structure and spreading dynamics lead to
highly diversified spreading capabilities among individual nodes.
From the perspective of the severity of a spreading process, the
most influential spreaders are those resulting in a much larger final
infected proportion of the whole system when the spread of a disease
or a piece of information originates from them than from other
nodes. Centrality measures such as the degree~\cite{freeman1978},
betweenness~\cite{freeman1977}, closeness~\cite{sabi1966},
eigenvector centrality~\cite{bonacich2001} and $k$-shell
coreness~\cite{bolobas1984} have been used to identify the most
influential spreaders. The degree is the simplest measure. In
social networks, for a example, an individual with a large degree
has more direct contacts to other people and is thus likely to be
more influential than one with a small degree for transmitting
disease or information. Subsequent research indicates that the
core nodes as identified by the $k$-shell decomposition are the
most influential spreaders~\cite{kitsak2010identification}.
Algorithms based on other centrality measures have been proposed
to improve the accuracy of identifying influential
spreaders~\cite{pei2013spreading,chen2013identifying,ren2014iterative,lv2016vital}.
They include the neighborhood coreness~\cite{bae2014identifying},
improved eigenvector
centrality~\cite{martin2014localization,alvarez2015eigen},
H-index~\cite{hirsch2005index,lv2016hindex} and nonbacktracking
centrality~\cite{radicchi2016leveraging}.

Methods other than centrality-based algorithms have also been
proposed for predicting how influential a node can be in a spread.
For example, by counting the number of possible infection paths of
various lengths, the final infection range can be estimated for a
spread originated from any node~\cite{bauer2012identifying}. The
degree distribution of clusters of infected nodes after certain
transmission events leads to a node property called the expected
force, which can be applied to predict the spreading influence of
all nodes under different epidemiological
models~\cite{lawyer2015understanding}.
The dynamic-sensitive centrality is able to locate influential
nodes from both topological features and the dynamical parameters,
such as the infection and recovery rates in a
susceptible-infected-recovered (SIR) spreading
model~\cite{liu2016locating}. In the $k$-truss decomposition, which
is a triangle-based extension of $k$-shell decomposition, the maximal
$k$-truss subgraph contains the most influential
spreaders~\cite{malli2016locating}.

In most studies on identifying influential spreaders so far, the
networks are taken to be unweighted and undirected. Each edge is
treated to be equivalent in its function, as in the centralities
and ranking methods~\cite{lv2016vital}. However, edges in a
network could be quite different~\cite{grady2012robust}. In
weighted networks, the weight of an edge reflects the strength of
the interaction between the connected nodes, as in situations
concerning the number of communications, size of trade, intimacy
of friendship and frequency of cooperation,
etc.~\cite{opsahl2010node,garas2012kshell,eidsaa2013score}. In
addition, edges may not be equally important in keeping the
network robust~\cite{wu2011onion}. An example is the
small influence on network robustness in food web networks when
redundant links are removed~\cite{allesina2009functional}. In
terms of network functionality, differences among edges are also
observed. For example, removing redundant links has no effect on
network synchronization~\cite{zhang2014network} but closing
specific routes in air transportation networks can minimize the
spreading of a disease~\cite{chung2012impact}. To quantify the
weight of an edge, a class of measures relying on its importance
in the network structure have been defined~\cite{brockmann2013hidden}. For example, the edge
betweeness counts the number of shortest paths between any two
nodes that go through the edge and it can be regarded as the
weight of an edge~\cite{brandes2001faster}. Immunizing edges of
high betweenness was found to be effective in suppressing
epidemics~\cite{holme2002attack}, but deleting such edges in
scale-free networks would enhance the transmission efficiency
dramatically~\cite{zhang2007enhancing}. In global air
transportation network, the strength of an edge that reflects the
volume of passengers travelling between two airports was found to
correlate positively with the product of the degree of the
connected nodes. Thus, a measure
$w_{ij}=(k_ik_j)^\theta$~\cite{barrat2004arch} was introduced as
the weight of an edge. This measure has been adopted in many
works for distinguishing the importance among edges in
unweighted networks~\cite{wang2005general,tang2011efficient,zhang2014suppression}.

In the present work, we propose a better measure to quantify the
importance of an edge when spreading processes are concerned. As
spreading is necessarily directional, e.g. only an infected node
would spread a disease to a neighboring susceptible node but not
the other way round, the new measure stresses the importance of an
edge in the spreading dynamics in the vicinity of the two nodes
connected by the edge and it has the general property of $w_{ij}
\neq w_{ji}$. The sum of asymmetric weights of links originated from a node
defines a new measure of the
strength $s_{i}$ of a node $i$, which is shown to be an efficient
quantity for identifying influential spreaders with a low
computational complexity. Based on the node strength, a
$s$-shell decomposition scheme is proposed for assigning a $s$-shell index
to every node, which provides a more accurate
ranking of the nodes in their influence in spreading processes.

The paper is organized as follows. In
Sec.~\ref{sec:centrality}, the degree centrality, $k$-shell index,
the spreading model used in the study and the methods of evaluating
the performance of measures for identifying influential spreaders are
introduced for completeness. In Sec.~\ref{sec:define}, we propose
a new measure that focuses on the importance of an edge in the
dynamics of a spreading process. The measure is then applied to
define a node strength for every node. A $s$-shell decomposition
method that emphasizes the importance of a node in the spreading dynamics is
proposed. In Sec.~\ref{sec:effect} we apply the node strength and $s$-shell index to
rank and identify influential spreaders in nine real-world networks and demonstrate
their effectiveness. A conclusion is given in
Sec.~\ref{sec:discussion}.

\section{Centralities, spreading model and evaluation methods} \label{sec:centrality}

We review briefly the degree centrality and the $k$-shell index
for completeness. They are efficient measures for identify
influential
spreaders~\cite{castellano2012competing,liu2015core,liu2015improving}.
We will compare the performance of our newly defined node strength and $s$-shell index with these methods. The
Susceptible-Infected-Recovered (SIR) model is adopted to simulate
the spreading dynamics on networks. To quantify the performance
of our measures in predicting the influence of the nodes and
identifying influential spreaders, the Kendall's correlation and the
imprecision function are introduced.

\subsection{The degree and $k$-shell centrality}
In a graph $G=(V,E)$, where $V$ is the set of nodes and $E$ is the
set of edges, the degree $k_{i}$ of a node $i$ is the number of
links it carries.  It is given by $k_i=\sum_{j} a_{ij}$, where
$a_{ij}$ is an element of the adjacent matrix, with $a_{ij}=1$ if
there is a link between nodes $i$ and $j$ and $a_{ij}=0$
otherwise. The $k$-shell decomposition method decomposes the
network into hierarchical shells in a progreesive process.
Initially, nodes with degree $k=1$ are removed from the network
together with their links. After the process, there may appear
nodes with only one link left. These nodes and their links are then
removed and the process is repeated until there is no nodes left
in the network with only one link. The removed nodes and links
form the $1$-shell, and these nodes are assigned with an index $k_S
= 1$. Next, nodes with degree $k\leq 2$ are removed in a similar
way and the set of removed nodes are assigned an index
$k_S=2$. This pruning process is continued until all nodes are
removed and assigned a $k_S$ index. This index is called the
$k$-shell index or coreness of a node. It represents the core
position of a node in the network. Nodes with a large $k_S$ are
considered as to be at the core of the network, while nodes with
a small $k_S$ form the peripheral part of the network.

Nodes with large degree and large coreness are considered the most
influential spreaders in networks. These measures have a low
computational complexity of order $O(E)$ and $O(N+E)$, where $N$ and $E$
are the number of nodes and edges in the network respectively.

\subsection{SIR model}
The SIR model is chosen to simulate spreading on complex networks.
In the model, the nodes have three possible states:
S (susceptible), I (infected) and R (recovered). At each time step,
the infected nodes infect their susceptible neighbors with a
probability $\lambda$ and then recover with a probability $\beta$.
To quantify the influence of each node on spreading, we let one
node, node $i$ say, be infected and all the other nodes being
susceptible initially. The SIR dynamics proceeds from the seed
infected node to other nodes until there is no infected node in
the network. The recovered nodes at the end are those once
infected and the fraction of recovered nodes gives the final
infected range of the initial seed. For an initially infected
node $i$, the spreading dynamics is repeated for 100 times. The
average infected range $M_{i}$ of node $i$ is recorded and taken
to reflect the influence or the spreading efficiency of the node $i$.
This quantity can be obtained for any node $i$ in the network and
used as a measure to rank the nodes on their importance in the
spreading dynamics. This dynamics-based list is taken to be the
exact ranking that gauges the accuracy of other topology-based
measures.

While the final infected ranges for the nodes vary with the
parameters $\lambda$ and $\beta$ in the SIR model, the relative
ranking of spreading efficiency of the nodes remains unchanged in
a wide range of infection probabilities~\cite{liu2015core}. Thus,
we take the recovered probability to be $\beta=1$ for simplicity.
The infection probability $\lambda$ should be chosen more
carefully. Too large an infection probability gives spreading
efficiencies of the nodes that are too close to each other to
distinguish their relative importance clearly. In the results that
follow, we choose an infection probability $\lambda$ above the
epidemic threshold that gives a final infected range that amounts
to $1\%$ to $20\%$ of the system for most
nodes~\cite{kitsak2010identification}.

\subsection{The Kendall's tau correlation and imprecision
function}
Two figures of merit are used to quantify the performance of
different topology-based measures for predicting the spreading
efficiency of the nodes. The Kendall's tau correlation
coefficient measures the ranking consistency of two lists that
rank the same set of objects. By referring to the number of
concordant ranking pairs and the number of discordant ranking
pairs in two ranking lists of $N$ objects, the correlation
coefficient is evaluated by
\begin{equation}
\tau=\frac{\sum_{i<j}sgn[(x_{i}-x_{j})(y_{i}-y_{j})]}{\frac{1}{2}N(N-1)}
\;,
\end{equation}
where sgn(x) is the sign function which returns $1$ if $x>0$, $-1$
if $x<0$, and 0 if $x=0$, and the summation is over all
distinguished pairs $i$ and $j$. Here, $x_{i}$ is the rank of node
$i$ in ranking list 1, while $y_{i}$ is the rank of node $i$ in
ranking list 2. In the present context, list 1 is a topology-based
ranking and list 2 is the SIR dynamics-based ranking. If $(x_{i} -
x_{j})$ has the same sign as $(y_{i} - y_{j})$, the two lists give
the same relative ranking of node $i$ and node $j$. Therefore, a
large $\tau$ implies a more concordant relation between two
methods of ranking the nodes.

For spreading processes, it is also important to quantify the
accuracy in pinpointing the most influential spreaders. For a
topology-based measure $\theta$, e.g. some kind of node
centrality, let $M_{\theta}(p)$ be the average spreading
efficiency of the $pN$ nodes carrying the highest measure
$\theta$. Similarly, let $M_{eff}(p)$ be the average spreading
efficiency of $pN$ nodes carrying the highest actual spreading
efficiency according to the SIR dynamics. The imprecision
function~\cite{kitsak2010identification}
\begin{equation}\label{imprecision}
\varepsilon_{\theta}(p)=1-\frac{M_{\theta}(p)}{M_{eff}(p)}
\end{equation}
quantifies how close is the average spreading of the $pN$ nodes
based on centrality ranking to the actual spreading. A smaller
$\varepsilon_{\theta}$ represents a higher accuracy of $\theta$ in identifying
the most influential spreaders.

\section{Dynamical importance of edges and weighted node centrality} \label{sec:define}
The dynamical importance of an edge is analyzed by focusing on the
spreading dynamics and the edge's local structure. This leads to
the necessity of assigning bi-directional and asymmetric weights
to an edge. A novel node strength $s$ can then be defined to
quantify the impact of a node on spreading. A $s$-shell
decomposition method is proposed to be a reliable way of ranking
the nodes for spreading processes.

\subsection{Dynamical importance of edges}
\begin{figure}
\begin{center}
\epsfig{file=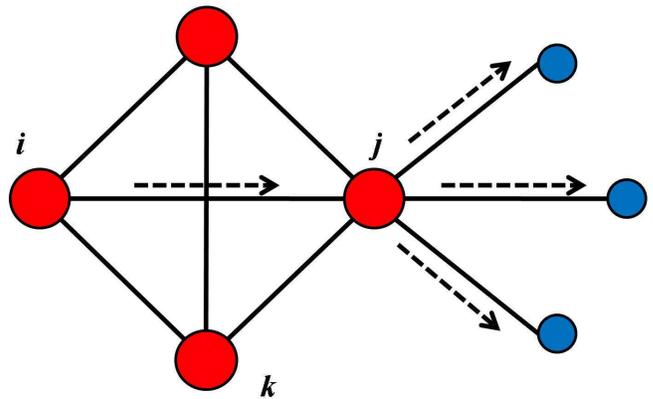,width=1\linewidth}
\caption{(Color online) {\bf Local structure of a network emphasizing the role of the
link $e_{ij}$ in spreading a disease from node $i$ to node $j$ and
then to reach out to nodes that node $i$ itself cannot reach.} The
same link $e_{ji}$, however, plays a different role as it does not
help spread the disease to nodes beyond the reach of node $j$
after it infects node $i$. The asymmetry requires the assignment
of directional weights with $w_{ij} \neq w_{ji}$.}\label{figure1}
\end{center}
\end{figure}

Figure~\ref{figure1} shows part of a network. When a disease
originates from node $i$ and spreads along the edge $e_{ij}$, node
$j$ will be infected first. Once node $j$ is infected, it could
spread to other parts of the network through node $j$'s
``out-reaching" edges, which are edges that connect node $j$ to
nodes that are not in $i$'s neighborhood. The number of
out-reaching edges from $j$ is denoted by $k_j^{out}$ and it is
three in the example of Fig.~\ref{figure1}. Note that
$k_{j}^{out}$ should depend on the node $i$, as $j$ must be a
neighboring node of $i$. In contrast, the edge $e_{ik}$ has zero
out-reaching edge after it is infected by node $i$. Therefore, the
edge $e_{ij}$ is expected to be more important in that it is more
likely to lead to a larger infected area than confining the
infection to node $i$'s neighborhood as $e_{ik}$ does
~\cite{lawyer2015understanding,liu2015improving}. We are,
therefore, motivated to introduce a new measure to distinguish the
different importance of edges in a spreading process, even though
the links may be unweighted in the construction of the network.

For our purpose, we define a weight $w_{ij}$ for an edge $e_{ij}$
by
\begin{equation}\label{1}
w_{ij}=1+(k_ik_j^{out})^a
\end{equation}
to represent its importance in a spreading process from node $i$
to node $j$. The first term stands for a basic and symmetric
effect of an edge. The factor $k_j^{out}$ is included to reflect
the potential impact of node $i$ through infecting its neighbor
$j$. The product $k_i k_j^{out}$ include the degree of node $i$
into consideration. The idea can be illustrated by considering a
leaf node (node of degree 1) connected to a hub (a node of large
degree), the number of out-reaching links is very large for this
leaf node. However, its impact is not necessarily high because
only when itself and its neighboring hub are infected that the
infection could spread to the other part of the network. The
parameter $a$ serves to tune the contribution of $k_ik_j^{out}$ to
the importance of edge $e_{ij}$.

The presence of $(k_{i}k_{j}^{out})^{a}$ emphasizes the {\em
asymmetric} importance of an edge. The weight $w_{ij}$ is
different from $w_{ji}$ for the same link connecting node $i$ and
node $j$. From Eq.~(\ref{1}), $w_{ji}=1+(k_jk_i^{out})^a$, which
measures the importance of the edge $e_{ji}$ when a spread goes
from node $j$ to node $i$ along the link $e_{ji}$ and then move on
to other parts of the network. Note that $w_{ij} \neq w_{ji}$
generally as they are defined by considering the neighborhoods of
the neighbors of node $i$ and node $j$, respectively. Given a
network, $w_{ij}$ and $w_{ji}$ can be evaluated entirely based on
the network topology and they label every edge to better reflect
the bidirectional and yet asymmetric contributions of the edge in
spreading processes.

\subsection{A novel node strength and $s$-shell decomposition}

It will be advantageous to introduce a node-level quantity
analogous to the degree to quantify the importance of a node in
spreading dynamics. This will put the computational complexity at
the same level as those based on the degree and $k$-shell
decomposition. Motivated by the idea of weighted
degree~\cite{opsahl2010node} that the strength of a node in a
weighted graph is the sum of the weights of its edges, we define
the strength $s_{i}$ of a node $i$ by
\begin{equation}\label{2} s_i=\sum_{j\in\Gamma_i}w_{ij},
\end{equation}
where the summation is over the nodes $j$ belonging to the
neighborhood $\Gamma_i$ of node $i$. Invoking $w_{ij}$ in the
definition of $s_{i}$ makes it a better measure in quantifying a
node's importance in spreading dynamics.

We propose a $s$-shell decomposition method as an extension of the
$k$-shell decomposition. The algorithm is as follows. With the
strengths $s_{i}$ evaluated for all nodes, the algorithm starts
with removing the nodes with the smallest strength $s_m$ and the
links associated with the nodes. Let node $i$ be removed. The
strength of its neighboring node $j$ is then updated to $s_{j} -
w_{ji}$ as the edge $e_{ij}$ is removed. The network is then
checked and the removal of nodes with $s_{m}$ continues until no
nodes of strength less than or equal to $s_{m}$ remains. The
deleted nodes are assigned a $s$-shell index of $s_s=1$, where the
symbol emphasizes that the decomposition is based on the nodes'
strength and the subscript represents a shell. The trimming
process is repeated for the nodes with the smallest strength among
the remaining nodes and the nodes so removed are assigned the
index $s_{s}=2$. This pruning process is continued until all nodes
are removed and assigned a $s_{s}$ index. The $s$-shell index of
a node can be regarded a weighted coreness of the node emphasizing
its importance in spreading dynamics.

\section{Performance in identifying influential spreaders in real-world networks} \label{sec:effect}

To examine the effectiveness of using the node strength and
weighted coreness in identifying influential spreaders, we apply
the measures to nine real-world networks as listed in
Table~\ref{tab:basiccharacteristic}. 
\begin{table*}
\begin{center}
\caption{Properties of the real-world networks studied in
this work. Structural properties of the number of nodes ($N$),
number of edges ($E$), average degree ($\langle k \rangle$),
degree assortativity ($r$), clustering coefficient ($C$), epidemic
threshold ($\lambda_c$), infection probability ($\lambda$) used in
the SIR dynamics, and the optimal value of $a$ as given by the
Kendall's tau correlation coefficient ($a_{opt}$).}
\begin{tabular}{ccccccccccc}
\hline
\textbf{Network} & \textbf{$N$} & \textbf{$E$} & \textbf{$\langle k \rangle$} & \textbf{$r$} & \textbf{$C$} & \textbf{$\lambda_c$} & \textbf{$\lambda$} & \textbf{$a_{opt}$}\\
\hline
CA-Hep &8638 &24806 &5.7 &0.239 &0.482 &0.08 &0.12 &1.0\\
Astro  &14845 &119652 &16.1 &0.228 &0.670 &0.02 &0.05 &0.9\\
Emailcontact  &12625 &20362 &3.2 &-0.387 &0.109 &0.01 &0.10 &0.3\\
PGP  &10680 &24340 &4.6 &0.240 &0.266 &0.06 &0.19 &1.0\\
Blog  &3982 &6803 &3.4 &-0.133 &0.284 &0.08 &0.27 &0.9\\
AS  &22963 &48436 &4.2 &-0.198 &0.230  &0.004 &0.13 &0.2\\
Router  &5022 &6258 &2.5 &-0.138 &0.012 &0.08 &0.27 &0.7\\
Hamster  &2000 &16097 &16.1 &0.023 &0.540 &0.02 &0.04 &0.8\\
Netsci  &379 &914 &4.8 &-0.082 &0.741 &0.14 &0.30 &0.8\\
\hline
\end{tabular}
\label{tab:basiccharacteristic}
\end{center}
\end{table*}

The real networks studied are: (1) CA-Hep (giant connected
component of collaboration network of arXiv in high-energy physics
theory)~\cite{leskovec2012}; (2) Astro physics (collaboration
network of astrophysics scientists)~\cite{newman2001};
(3) Emailcontact (email contacts at Computer Science Department of
University College London)~\cite{kitsak2010identification}; (4)
PGP (an encrypted communication network)~\cite{boguna2004}; (5)
Blog (the communication relationships between owners of blogs on
the MSN (Windows Live) Spaces website)~\cite{xie2006}; (6) AS
(Internet at the autonomous system level)~\cite{newmandataas}; (7)
Router (the router level topology of the Internet, collected by
the Rocketfuel Project)~\cite{spring2004}; (8) Hamster
(friendships and family links between users of the website
hamsterster.com)~\cite{hamster2014}; and (9) Netsci (collaboration
network of network scientists)~\cite{newman2006}. The new
measures are found to outperform predictions based on the degree
centrality and $k$-shell decomposition, as we now show.

\subsection{Performance of node strength}

\begin{figure}
\begin{center}
\epsfig{file=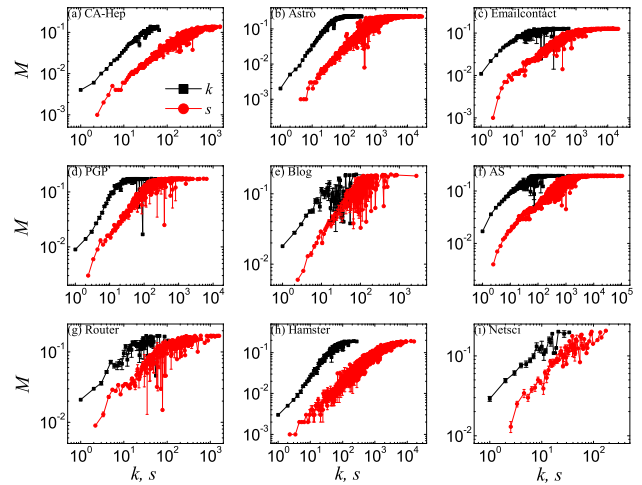,width=1\linewidth}
\caption{(Color online) {\bf Mean spreading efficiency $M$ of nodes as classified by
their degree $k$ (black squares) or their node strength s (red
circles) in nine real-world networks.} As the node strength are
real numbers instead of integers, data are grouped in intervals of
size unity. The corresponding average spreading efficiency and
node strength in each interval are displayed, starting from the
minimal strength.}\label{figure2}
\end{center}
\end{figure}

From the structure of each network, every node carries a degree
$k_{i}$ and a node strength $s_{i}$. Using the SIR dynamics, the
spreading efficiency $M_{i}$ of each node can be obtained by
simulations. Fig.~\ref{figure2} compares the correlations between
the spreading efficiency with the node strength and with the
degree in nine real-world networks. Here, we take
$a=0.5$ in Eq.~(\ref{1}) in determining $w_{ij}$ for the edges.
The sensitivity to the parameter $a$ will be discussed later.
The strength and the degree are both positively correlated with
the spreading efficiency. The merit of using the strength over
the degree as a measure is that their values cover a much wider
range and they can distinguish the spreading efficiency more
specifically. This advantage is built into the definition of the
node strength as it captures the key elements in spreading
dynamics.

\begin{figure}
\begin{center}
\epsfig{file=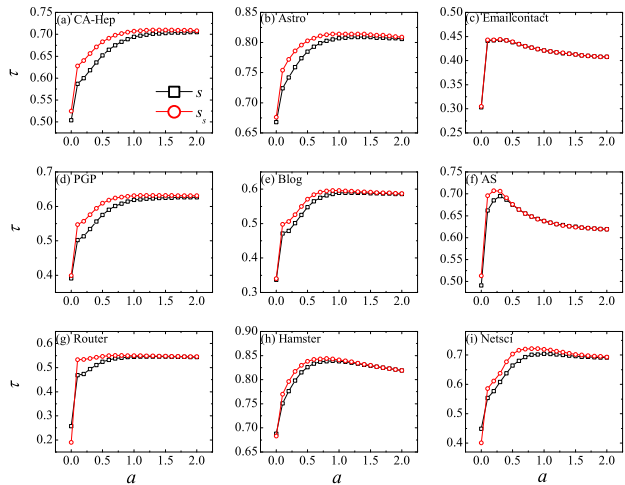,width=1\linewidth}
\caption{(Color online) {\bf Kendall's tau correlation coefficients evaluated between
the actual spreading efficiency of the nodes and the ranking based
on node strength (black square) and based on the $s$-shell
index (red circle) for different values of the parameter $a$ in
Eq.~(\ref{1}).}}\label{figure3}
\end{center}
\end{figure}

The node strength provide a ranking of the nodes. This list can
be compared with the list based on the actual spreading efficiency
by calculating the Kendall's tau correlation coefficient. We
calculate the ranking correlation of nodes' spreading efficiency and their strength for different
values of $a$ and obtained $\tau(a)$, as shown in
Fig.~\ref{figure3} (squares).
For $a=0$ (see Eq.~(\ref{2})), $s_{i}$ reduces to the degree
$k_{i}$ and thus $\tau (a=0)$ measures the correlation between the
rankings based on the degree and the spreading efficiency. Note
that $\tau$ is significantly enhanced for $a>0$, implying that the
node strength, which includes the bi-directional and asymmetric
weights of the edges, ranks the nodes more accurately. Results in
Fig.~\ref{figure3} further show that there exists an optimal value
of $a$ for each network at which $\tau$ is a maximum. The optimal
value of each network is given in Table 1, together with the other
network properties. Fig.~\ref{figure3} also shows the $\tau(a)$
obtained by ranking the nodes according to the $s$-shell index.
The results will be discussed later.

\begin{figure}
\begin{center}
\epsfig{file=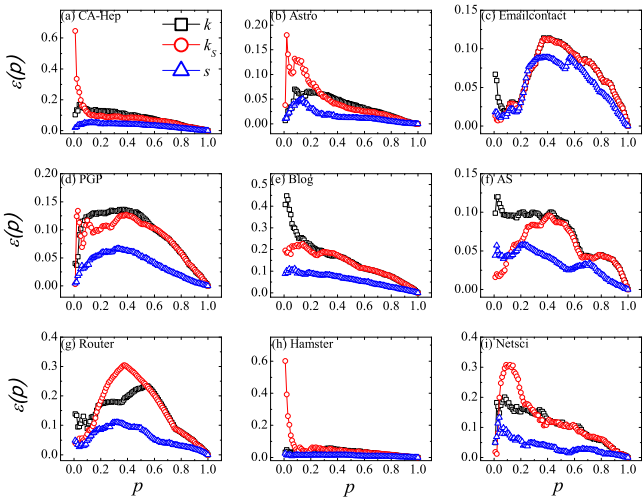,width=1\linewidth}
\caption{(Color online) {\bf The imprecision of rankings based on the degree ($k$,
black squares), $k$-shell index ($k_S$,red circles) and node strength ($s$, blue triangles) evaluated at the optimal value of
$a$ as a function of $p$ for nine real-world networks.} The node strength provide a better measure for identifying influential
spreaders.}\label{figure4}
\end{center}
\end{figure}

Fig.~\ref{figure4} shows the imprecision function of the ranking
based on the node strength, together with the results based on
the degree and $k$-shell index for comparison.
Recall that a lower imprecision implies a higher accuracy in
identifying the influential spreaders. The node strength
(triangles) give an imprecision which is less than 0.1 for all $p$
in nearly all cases. Only in the network Netsci, the imprecision
is slightly larger than 0.1 for a few values of $p$. The results
show that the node strength outperforms the degree (squares) in
accuracy in almost all networks. Only in the network Hamster, the
imprecisions based on node strength and on the degree become
comparable but they are both small. The node strength is,
therefore, a better index for pinpointing the influential
spreaders than the degree. More noticeably is that the node
strength performs even better than the $k$-shell index in most
cases, except at some small values of $p$ in the AS and Netsci
networks. The $k$-shell index is regarded as an efficient measure
for identifying influential spreaders and it is widely used in
ranking algorithms. However, the assignment of $k$-shell index
requires a higher computational complexity and complete network structure than index relying solely on
node-level quantities such as the degree or the node strength.
The node strength introduced here does not only provide a more
accurate measure, but also a computationally efficient method in
handling large-scale networks.

\subsection{Performance of weighted coreness}
The $k$-shell index works better than the degree in identifying
influential spreaders~\cite{liu2015improving}. Here, we investigate how the $s$-shell
index $s_{s}$ or weighted coreness works in comparison to the
other measures. The results of the Kendall's tau correlation of
$s_{s}$ ranking in Fig.~\ref{figure3} suggest that it is a better
measure than using the node strength in eight systems out of nine.
In the Emailcontact network, $s_{s}$ and $s$ rankings work equally
well. In fact, the $s_{s}$ and $s$ rankings approach the same
value of $\tau$ as $a$ increases. Given that the optimal values
of $a$ in the networks are less than or equal to $1$, the weighted
coreness gives a better ranking. Note that the $a=0$ case gives
the value of $\tau$ corresponding to the  $k$-shell index $k_{S}$.
Using $s_{s}$ to rank the nodes always gives a higher $\tau$ than
the $a=0$ value, implying $s$-shell index is also a better measure
than the $k$-shell index.

\begin{figure}
\begin{center}
\epsfig{file=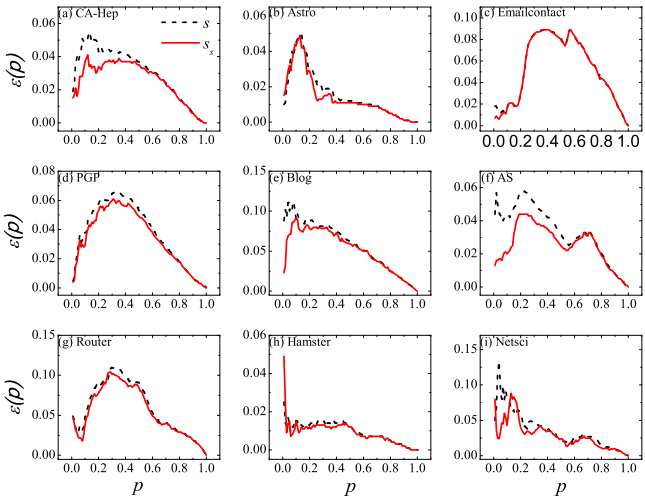,width=1\linewidth}
\caption{(Color online) {\bf The imprecision of rankings based on the node strength
($s$, black dash) and weighted coreness ($s_s$, red solid)
obtained by $s$-shell decomposition, as a function of $p$.} The
weighted coreness $s_s$ provides a further improvement over $s$ in
identifying influential spreaders.}\label{figure5}
\end{center}
\end{figure}

The imprecision functions of rankings using $s_{s}$ and $s$ are
compared in Fig.~\ref{figure5}. Their performances are comparable
and they both work better than measures based on the degree alone
(see Fig.~\ref{figure4}). Looking closer, the lower imprecision
of $s_{s}$ ranking in six (CA-Hep, PGP, Blog, AS, Router and
Netsci) out of nine cases suggests that the $s$-shell
decomposition method is more accurate in identifying influential
spreaders in real-world networks. Even in the networks of Astro,
Emailcontact and Hamster that $s_{s}$ and $s$ work almost equally
well, the imprecision of $s_s$ is slightly lower or equal to that
of $s$. Only in the Hamster network that $s$ works slightly better
than $s_{s}$ at $p=0.01$, even so the imprecision functions are
small (under 0.05) on the absolute scale.

\subsection{Robustness of proposed weighted centrality}

\begin{figure}
\begin{center}
\epsfig{file=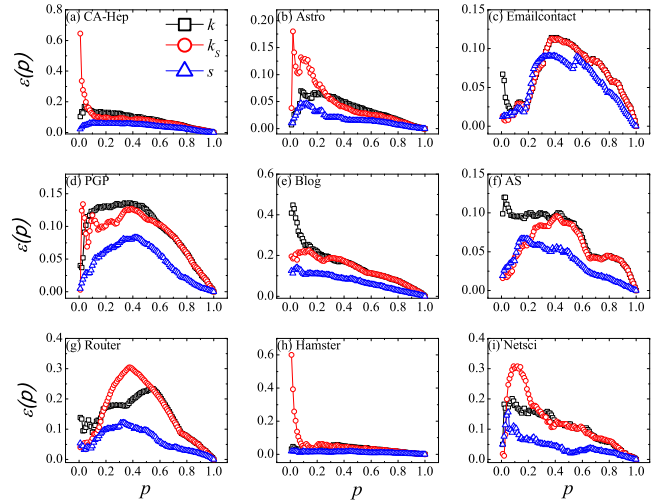,width=1\linewidth}
\caption{(Color online) {\bf The imprecision of rankings based on the degree ($k$,
black squares), $k$-shell index ($k_S$,red circles) and node strength ($s$, blue triangles) evaluated by $a=0.5$ for nine
real-world networks as a function of $p$.}}\label{figure6}
\end{center}
\end{figure}
So far, we have used the optimal value of $a$ to evaluate $w_{ij}$ and $s_{i}$, and compared results with other measures. However,
the optimal value is not often known precisely in real
applications. It will be useful to examine the performance of the
node strength $s_{i}$ for some arbitrarily chosen value of $a$.
Let us set $a=1/2$ so that the term $(k_{i}k_{j}^{out})^{a}$ in
$w_{ij}$ represents a geometric mean. The comparison in
Fig.~\ref{figure6} of the imprecision function shows that $s_{i}$ ranking gives a lower imprecision than the degree and $k$-shell index. An interesting point is that
the imprecision of node strength evaluated at $a=1/2$ is even
lower than that evaluated at the optimal value of $a$ in the AS
network for $p<0.1$. The result indicates that although the best
performing overall ranking correlation coefficient occurs at some
optimal $a$, the same value does not necessarily give the best
identification of the most influential spreaders.

\begin{figure}
\begin{center}
\epsfig{file=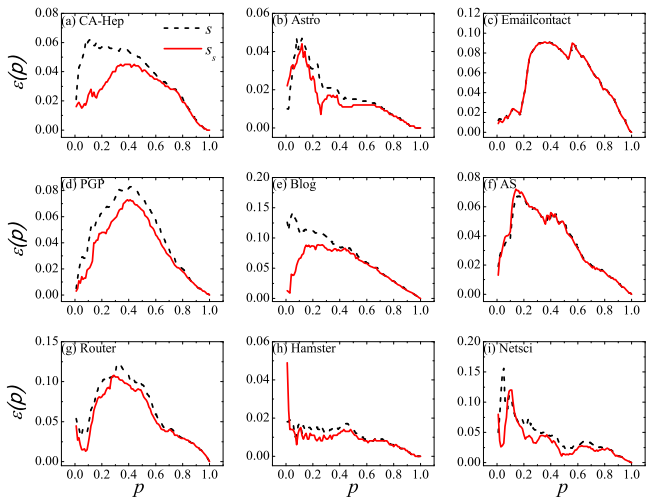,width=1\linewidth}
\caption{(Color online) {\bf The imprecision of rankings based on the node strength
($s$, black dash) and weighted coreness ($s_s$, red solid)
obtained by $s$-shell decomposition for $a=1/2$ as a function of
$p$.}}\label{figure7}
\end{center}
\end{figure}

Fig.~\ref{figure7} compares the effectiveness of the node strength and the $s$-shell decomposition for $a=1/2$. Again, the
$s$-shell index works better in most cases. In fact, the results
resemble those in Fig.~\ref{figure5} when the optimal $a$ is used.
These results further support the assertion that the node strength and the corresponding $s$-shell index are better
measures for spreading processes than methods based on the
degree. Between them, the $s$-shell index performs slightly
better, but evaluating the index requires more computing effort
than the node strength alone.

\section{Conclusion} \label{sec:discussion}
The roles of nodes and edges in deciding the structural properties
of a network should be carefully distinguished from their roles in
determining the extent of spreading processes. Although an edge
between nodes $i$ and $j$ certainly helps spread a disease, its
role may be different when the infection goes from $i$ to $j$ than
in the other direction. It is because what matters is whether the
node $j$, after infected by $i$, would reach out to other nodes
that node $i$ itself could not reach. If so, the link carries a
greater importance for infection from $i$ to $j$ which is
quantified by a higher weight $w_{ij}$ for the link. It is,
therefore, necessary to invoke asymmetric and bidirectional
weights with $w_{ij} \neq w_{ji}$ for a link so as to capture the
dynamics in spreading processes. Here, we introduced a form of
$w_{ij}$ (see Eq.~(\ref{1})) and showed that it facilitates more
accuracy ranking in the node's importance. Pictorially, the
network is better described by the nodes connected by links with
asymmetric weights in different directions when spreading dynamics
is concerned.

To establish the effectiveness of our method, the weights of the
links were used to construct a node strength $s$ that predicts the
importance of a node in spreading processes. A $s$-shell
decomposition scheme based on the node strength was then
introduced. The $s$-shell index $s_{s}$ of the nodes provide
another way of ranking them. Applying $s$ and $s_{s}$ rankings to
nine real-world networks, it was found that our novel measures
generally outperform the standard rankings based on the degree of
the nodes and the $k$-shell decomposition method. Superiority is
shown in both the overall performance of the ranking as indicated
by Kendall's tau correction coefficient and in identifying the
influential spreaders as indicated by the imprecision.

\begin{figure}
\begin{center}
\epsfig{file=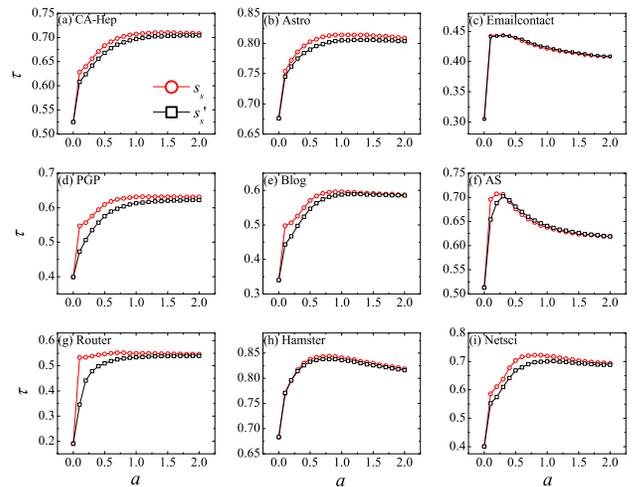,width=1\linewidth}
\caption{(Color online) {\bf Kendall's tau correlation coefficients evaluated between
the actual spreading efficiency of the nodes and the rankings
based on weighted coreness $s_{s}$ (red circles) as obtained by
assigning asymmetric weighting $w_{ij}$ of Eq.~(\ref{1}) to the
links and based on weighted coreness $s_{s}'$ (black square) by
assigning symmetric weighting $w_{ij}'$ of Eq.~(\ref{3}) to the
links.} The necessity of invoking asymmetric weights is
demonstrated by the higher accuracy in the ranking based on $s_s$
than $s_{s}'$.}\label{figure8}
\end{center}
\end{figure}

The success of our measure relies on the asymmetry in the weights
contained in $w_{ij}$ and $w_{ji}$. To stress the point, we
constructed a related network with weighted links but the weights
are symmetric by assigning a weight $w_{ij}'$ to a link according
to
\begin{equation}\label{3}
w_{ij}'=\frac{1}{2}(w_{ij}+w_{ji}) \;,
\end{equation}
with $w_{ij}$ given by Eq.~(\ref{1}). The weights $w_{ij}'$ can
then be used to assign a strength $s'$ to the nodes and a
corresponding $s$-shell decomposition based on $s'$ can be carried
out to assign an index $s_{s}'$ to each node. Fig.~\ref{figure8}
compares the Kendall's tau correlation of rankings based on
$s_{s}$ and $s_{s}'$ with the actual SIR spreading efficiency for
different values of the parameter $a$. In all cases, the measure
with asymmetric weights $s_{s}$ works better than that without the
asymmetry. In the Emailcontact and Hamster networks, the two
measures are equally accurate. The results confirm that it is
important to include the different roles of a link in spreading a
disease between nodes $i$ and $j$ in two different directions into
the construction of a reliable measure.

In summary, we proposed a node strength as an alternative
centrality measure for efficient and accurate identification of
influential spreaders. The idea of examining the functionality of
a link in spreading in either directions is a general one and thus
could be further developed for ranking a set of objects. We used
the SIR model as the spreading dynamics. However, the idea of
invoking asymmetric weights $w_{ij} \neq w_{ji}$ for a link
remains valid for other processes such as rumor spreading and
information diffusion, although the exact form of the weights may
depend on the details of the process under consideration.

\acknowledgments

This work was jointly funded by the National Natural Science
Foundation of China (Grant No. 11575041, 61672238), the Scientific Research
Starting Program of Southwest Petroleum University (No.
2014QHZ024), the Data Intelligence Academic Innovation Team of
Southwest Petroleum University (2015CXTD06) and the Fundamental
Research Funds for the Central Universities (Grant No.
ZYGX2015J153).


\begin{thebibliography}{100}
\bibitem{freeman1978}
L. C. Freeman, Centrality in social networks conceptual clarification,
Soc. Net. \textbf{1} 215 (1978).

\bibitem{freeman1977}
L. C. Freeman, A set of measures of centrality based upon betweenness,
Sociometry \textbf{40} 35 (1977).

\bibitem{sabi1966}
G. Sabidussi, The centrality index of a graph, Psychometrika
\textbf{31} 581 (1966).

\bibitem{bonacich2001}
 P. Bonacich and P. Floyd, Eigenvector-like measures of centrality for
asymmetric relations, Soc. Net. \textbf{23} 191 (2001).

\bibitem{bolobas1984}
B. Bolob{\'a}s, Graph Theory and Combinatorics: Proc. Cambridge
Combinatorial Conf. in honor of P. Erd\"{o}s, Academic Press
NY (1984).

\bibitem{kitsak2010identification}
M. Kitsak, L. K. Gallos, F. Havlin, L. Liljeros, H. E. Muchnik, H. E. Stanlney and H. A. Makse, Identification of influential spreaders in
complex networks, Nat. Phys. \textbf{6} 888 (2010).

\bibitem{pei2013spreading}
S. Pei and H. A. Makse, Spreading dynamics in complex networks, J.
Stat. Mech. \textbf{12} 12002 (2013).

\bibitem{chen2013identifying}
D.-B. Chen, H. Gao, L. L\"{u} and T. Zhou, Identifying influential nodes
in large-scale directed networks: The role of clustering, PLoS
ONE \textbf{8} e77455 (2013).

\bibitem{ren2014iterative}
Z.-M. Ren, A. Zeng, D.-B. Chen, H. Liao and J.-G. Liu, Iterative resource
allocation for ranking spreaders in complex networks, Europhys. Lett.
\textbf{106} 48005 (2014).

\bibitem{lv2016vital}
L. L\"{u}, D.-B. Chen, X.-L. Ren, Q.-M. Zhang, Y.-C. Zhang and T. Zhou, Vital
nodes identification in complex networks, Phys. Rep.
\textbf{650} 1 (2016).

\bibitem{bae2014identifying}
J. Bae and S. Kim, Identifying and ranking influential spreaders in
complex networks by neighborhood coreness Physica A
\textbf{395} 549 (2014).

\bibitem{martin2014localization}
T. Martin, X. Zhang and M. E. J. Newman, Localization and centrality in
networks, Phys. Rev. E \textbf{90} 052808 (2014).

\bibitem{alvarez2015eigen}
A. J. Alvarez-Socorro, G. C. Herrera-Almarza and
L. A. Gonz\'{a}lez-D\'{\i}az, Eigencentrality based on dissimilarity
measures reveals central nodes in complex networks Sci. Rep.
\textbf{5} 17095 (2015).

\bibitem{hirsch2005index}
J. E. Hirsch, An index to quantify an individual's scientific
research output Proc. Natl. Acad. Sci. USA \textbf{102}
16569 (2005).

\bibitem{lv2016hindex}
L. L\"{u}, T. Zhou, Q. M. Zhang and H. E. Stanlney, The H-index of a
network node and its relation to degree and coreness Nat.
Comm. \textbf{7} 10168 (2016).


\bibitem{radicchi2016leveraging}
F. Radicchi and C. Castellano, Leveraging percolation theory to
single out influential spreaders in networks Phys. Rev. E
\textbf{93} 062314 (2016).

\bibitem{bauer2012identifying}
F. Bauer and J. T. Lizier, Identifying influential spreaders and
efficiently estimating infection numbers in epidemic models: A
walk counting approach, Europhys. Lett. \textbf{99} 68007 (2012).

\bibitem{lawyer2015understanding}
G. Lawyer Understanding the influence of all nodes in a network,
Sci. Rep. \textbf{5} 8665 (2015).

\bibitem{liu2016locating}
J.-G. Liu, J.-H. Lin, Q. Guo and T. Zhou, Locating influential nodes via
dynamics-sensitive centrality Sci. Rep. \textbf{6} 21380 (2016).

\bibitem{malli2016locating}
F. D. Malliaros, M.-E. G. Rossi and M. Vazirgiannis, Locating influential
nodes in complex networks, Sci. Rep. \textbf{6} 19307 (2016).

\bibitem{grady2012robust}
D. Grady, C. Thiemann and D. Brockmann, Robust classification of
salient links in complex networks, Nat. Comm. \textbf{3} 864 (2012).

\bibitem{opsahl2010node}
T. Opsahl, F. Agneessens and J. Skvoretz, Node centrality in weighted
networks: Generalizing degree and shortest paths Soc. Net.
\textbf{32} 245 (2010).

\bibitem{garas2012kshell}
A. Garas, F. Schweitzer and S. Havlin, A $k$-shell decomposition
method for weighted networks New J. Phys. \textbf{14} 083030 (2012).

\bibitem{eidsaa2013score}
M. Eidsaa and E. Almaas, S-core decomposition: A generalization of
k-core analysis to weighted networks Phys. Rev. E \textbf{88}
062819 (2013).

\bibitem{wu2011onion}
Z.-X. Wu and P. Holme, Onion structure and network robustness,
Phys. Rev. E \textbf{84} 026106 (2011).

\bibitem{allesina2009functional}
S. Allesina, A. Bodini and M. Pascual, Functional links and robustness
in food webs, Phil. Trans. R. Soc. B \textbf{364} 1701 (2009).

\bibitem{zhang2014network}
C.-J. Zhang and A. Zeng, Network skeleton for synchronization:
identifying redundant connections, Physica A \textbf{402} 180 (2014).

\bibitem{chung2012impact}
N. N. Chung, L. Y. Chew, J. Zhou and C. H. Lai, Impact of edge removal on
the centrality betweenness of the best spreaders, Europhys. Lett.
\textbf{98} 58004 (2012).

\bibitem{brockmann2013hidden}
D. Brockmann and D. Helbing, The hidden geometry of complex network-driven contagion phenomena, Science
\textbf{342} 1337 (2013).

\bibitem{brandes2001faster}
U. A. Brandes faster algorithm for betweenness centrality, J.
Math. Sociology \textbf{25} 163 (2001).

\bibitem{holme2002attack}
P. Holme, B. J. Kim, C. N. Yoon and S. K. Han, Attack vulnerability of
complex networks, Phys. Rev. E \textbf{65} 056109 (2002).

\bibitem{zhang2007enhancing}
G.-Q. Zhang, D. Wang and G.-J. Li, Enhancing the transmission efficiency
by edge deletion in scale-free networks, Phys. Rev. E
\textbf{76} 017101 (2007).

\bibitem{barrat2004arch}
A. Barrat, M. Barth\'{e}lemy, R. Pastor-Satorras and A. Vespignani, The
architecture of complex weighted networks, Proc. Natl. Acad.
Sci. USA \textbf{101} 3747 (2004).

\bibitem{wang2005general}
W.-X. Wang, B.-H. Wang, B. Hu, G. Yan and Q. Ou, General dynamics of
topology and traffic on weighted technological networks, Phys.
Rev. Lett. \textbf{94} 188702 (2005).

\bibitem{tang2011efficient}
M. Tang and T. Zhou, Efficient routing strategies in scale-free
networks with limited bandwidth, Phys. Rev. E \textbf{84}
026116 (2011).

\bibitem{zhang2014suppression}
H. F. Zhang, J. R. Xie, M. Tang and Y. C. Lai, Suppression of epidemic
spreading in complex networks by local information based
behavioral responses, Chaos \textbf{24} 043106 (2014).

\bibitem{castellano2012competing}
C. Castellano and R. Pastor-Satorras, Competing activation mechanism
in epidemics on networks, Sci. Rep. \textbf{2} 371 (2012).

\bibitem{liu2015core}
Y. Liu, M. Tang, T. Zhou and Y. Do, Core-like groups result in
invalidation of identifying super-spreader by $k$-shell
decomposition, Sci. Rep. \textbf{5} 9602 (2015).

\bibitem{liu2015improving}
Y. Liu, M. Tang, T. Zhou and Y. Do, Improving the accuracy of the
$k$-shell method by removing redundant links: From a perspective
of spreading dynamics, Sci. Rep. \textbf{5} 13172 (2015).

\bibitem{liu2016identifying}
Y. Liu, M. Tang, T. Zhou and Y. Do, Identify influential spreaders in
complex networks, the role of neighborhood, Physica A
\textbf{452} 289 (2016).

\bibitem{leskovec2012}
J. Leskovec, J. Kleinberg and C. Faloutsos, Graph Evolution:
Densification and Shrinking Diameters, ACM Trans. on Knowledge
Discovery from Data (ACM TKDD) \textbf{1} 1 (2007).

\bibitem{newman2001}
M. E. J. Newman, The structure of scientific collaboration networks, Proc. Natl. Acad. Sci. USA \textbf{98} 404 (2001).

\bibitem{boguna2004}
M. Bogu{\~n}{\'a}, R. Pastor-Satorras, A. Diaz-Guilera and A. Arenas, A
Models of social networks based on social distance attachment, Phys. Rev. E \textbf{70} 056122 (2004).

\bibitem{xie2006}
N. Xie, Social network analysis of blogs, M.Sc. Dissertation,
University of Bristol \textbf{} (2006).

\bibitem{newmandataas}
M. E. J. Newman, Network data,  Available at:
http://www-personal.umich.edu/\%7Emejn/netdata (Accessed:
12/12/2012).

\bibitem{spring2004}
N. Spring, R. Mahajan, D. Wetherall and T. Anderson, Measuring ISP
topologies with Rocketfuel, IEEE/ACM Trans. Networking
\textbf{12} 2 (2004).

\bibitem{hamster2014}
J. Kunegis, Hamsterster full network dataset - KONECT
Available at:
http://konect.uni-koblenz.de/networks/petster-hamster (Accessed:
01/03/2014).

\bibitem{newman2006}
M. E. J. Newman, Finding community structure in networks using the
eigenvectors of matrices, Phys. Rev. E \textbf{74} 036104 (2006).



\end{thebibliography}

\end{document}